\documentclass[a4paper,11pt]{article}
\usepackage{amsmath,amssymb,amsfonts}
\usepackage{float}
\usepackage{graphicx}
\usepackage{array,booktabs,tabularx}
\pdfoutput=1 

\usepackage{jinstpub} 
\usepackage{lineno}

\title{Effect of passive metallic layers on muon energy estimation by means of deflection angle for muon scattering tomography: A comparative study based on GEANT4 simulations}

\author[a,b,1]{A. I. Topuz,\note{Corresponding author.}}
\author[a,c]{M. Kiisk,}
\author[b]{A. Giammaco,}
\author[c]{M. M\"{a}gi}

\affiliation[a]{Institute of Physics, University of Tartu, W. Ostwaldi 1, 50411, Tartu, Estonia}
\affiliation[b]{Centre for Cosmology, Particle Physics and Phenomenology, Universit\'e catholique de Louvain, Chemin du Cyclotron 2, B-1348 Louvain-la-Neuve, Belgium}
\affiliation[c]{GScan OU, Maealuse 2/1, 12618 Tallinn, Estonia}

\emailAdd{ahmet.ilker.topuz@ut.ee, ahmet.topuz@uclouvain.be}

\abstract{In the tomographic configurations based on the muon scattering, the angular variation with respect to the kinetic energy indirectly brings forth the ability to coarsely predict the kinetic energy by using the deflection angle owing to the detector layers. Nevertheless, the angular deviation due to the detector components is expected to be minuscule in addition to a relatively high uncertainty in the case of the plastic scintillators. In the present study, we contrast our current tomographic prototype, which consists of the detector layers manufactured from polyvinyl toluene besides a detector accuracy of 1 mrad, with an alternative hodoscope scheme containing stainless steel layers by aiming to investigate the three-group energy structure. Initially, we determine the average deflection angles together with the corresponding standard deviations for our present setup as well as for the alternative scheme by means of the GEANT4 simulations. In the second place, we express a brace of misclassification probabilities founded on the standard deviations where the first procedure assumes a linear finite approximation, whereas the latter approach rests on a positively defined modified Gaussian distribution. Upon our simulation results, we demonstrate that the introduced stainless steel layers in the proposed hodoscope setup do not only serve to augment the average deflection angles, but they also diminish the misclassification probabilities, therewith reducing the classification uncertainty apart from an improved detection performance. }


\arxivnumber{2109.10518}

\proceeding{22$^{\text{nd}}$ International Workshop on Radiation Imaging Detectors\\
 June 27, 2021 to July 1, 2021\\
 Online}

\begin{document}
\maketitle
\section{Introduction}
\label{sec:intro}
During the encounter between the primary cosmic rays and the earth atmosphere, a non-negligible number of muons are generated over a wide energy spectrum. The fundamental basis, on which the muon scattering tomography is founded, is to follow the propagation of the cosmic ray muons within the volume-of-interest (VOI) where the entering muons of a certain energy deflect from their initial directions in the wake of the physical processes primarily hinging on the atomic number, the material density, and the material thickness~\cite{Checchia_2016, procureur2018muon, bonechi2020atmospheric}. Among the detection modules existing in the tomographic setups based on the muon scattering are the plastic scintillators that have substantially found their application by accentuating their favorable aspects like fast rise and decay times, high optical transmission, ease of manufacturing, low cost, and large available size~\cite{lee2017characteristics}. The hodoscope structure for the scattering-based tomography consists of two sections that are installed atop and beneath the VOI under the investigation~\cite{borozdin2003radiographic}, and each section is composed of two or more distinct detector layers~\cite{thomay2016passive, frazao2019high} occasionally made out of plastic scintillators usually with a moderate thickness. 

Typical muon tomography systems are inherently incapable of directly measuring the kinetic energy of incoming muons, as that would demand the presence of a magnetic field or of Cherenkov detectors or of a very precise time-of-flight measurement, and all those options would make the cost of the apparatus increase by a large factor.
However, the capability of assessing, even roughly, the energy of the incoming muons on an event per event basis can significantly enhance the precision of the tomography~\cite{anghel2015plastic}, as the deflection process depends directly on energy. 
In the course of the muon propagation through the detection system, the hodoscope components slightly contribute to the deviation of the transversing muons up to a certain extent, and this tiny contribution might serve to categorize the detected muons by building a binary relation between the deflection angle and the muon energy~\cite{georgadze2021method}.

By recalling the intrinsic properties of the plastic scintillators such as low atomic number and low density apart from a low thickness on the basis of the deflection angle, the strategic introduction of the passive metallic layers is expected to numerically ameliorate the angular outcomes for the purpose of the energy estimation thanks to the deflection power of the inserted material, which leads to the visible reduction in the angular uncertainty as well as the drastic augmentation in the tiny angular values. 
In this study, we categorize the incoming muons into three groups based on their deflection angle within the hodoscope that we use as a proxy for their kinetic energy.
We compare the classification performance in this three-group energy structure between our current tomographic prototype that contains three detector layers manufactured from polyvinyl toluene with a thickness of 0.4 cm as well as an accuracy of 1 mrad in both the top section and the bottom section, and an alternative hodoscope scheme including stainless steel layers in addition. 
In order to quantify the uncertainty propagated through the energy groups, we present a couple of misclassification probabilities characterized by one standard deviation where the first means supposes a linear finite approximation in one dimension, while the latter procedure is governed by the positively defined modified Gaussian distributions in two dimensions. The present study is organized as follows. In section~\ref{sec:Angle}, we express the average deflection angle as well as the corresponding standard deviation in terms of the hit positions and we also introduce the corresponding expressions averaged over the top section as well as the bottom section to diminish the width of the deflection angle. We define two forms of misclassification probabilities based on the standard deviations in section~\ref{sec:Mis}. While we demonstrate our simulation setup in the GEANT4 code~\cite{agostinelli2003geant4} besides the simulation features in section~\ref{sec:Sim}, we exhibit our simulation results in section~\ref{sec:Out}. Finally, section~\ref{sec:Con} narrates our concluding remarks.
\section{Average deflection angle and standard deviation}
\label{sec:Angle}
\begin{figure}[H]
\begin{center}
\includegraphics[width=10cm]{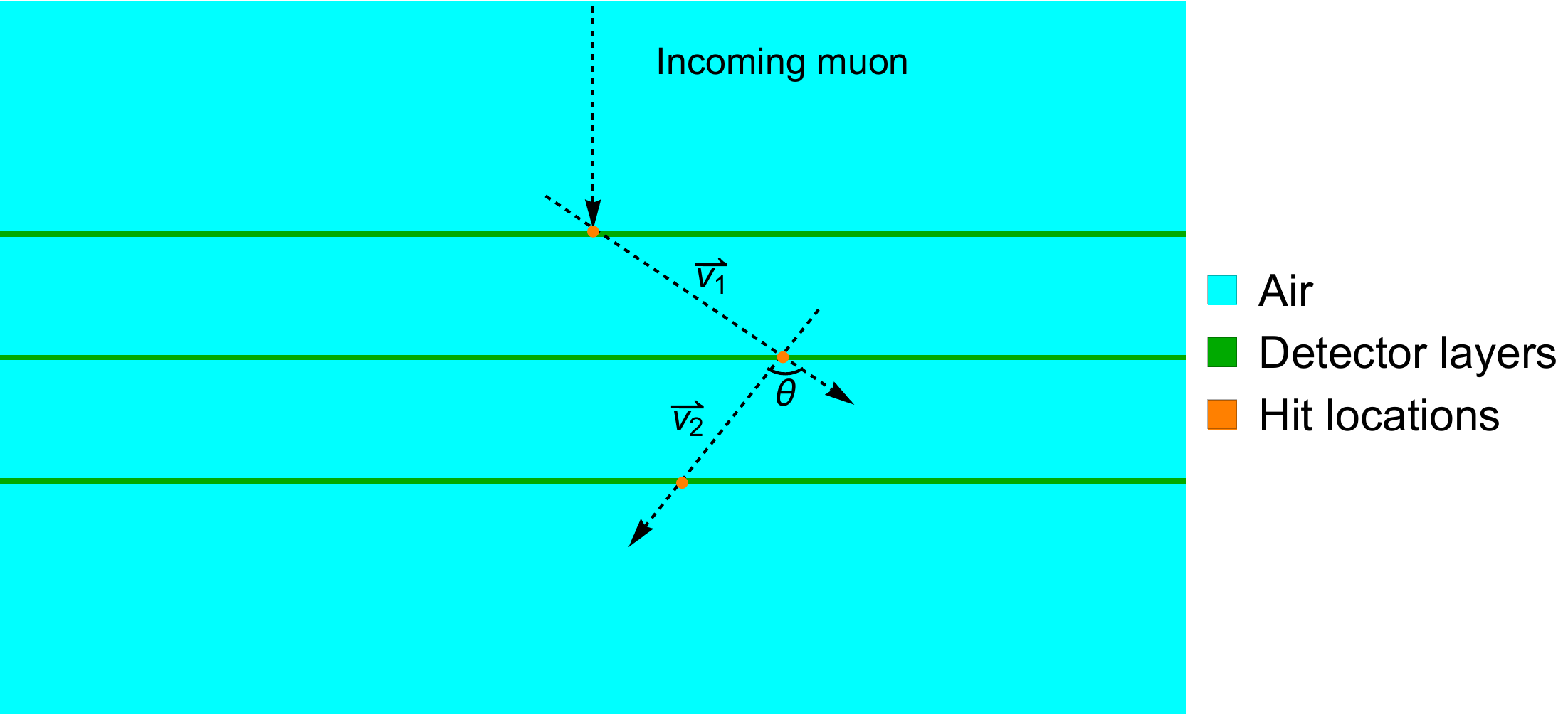}
\caption{Delineation of deflection angle denoted by $\theta$ via the hit locations on the detector layers made out of polyvinyl toluene.}
\label{Deflection_angle}
\end{center}
\end{figure}

By reminding that the crossing muons undergo deviations from the initial trajectory owing to the interactions with matter, the deflection angle of the incoming muons throughout a set of three detector layers is estimated via constructing two distinct vectors where the first vector is defined in accordance with the hit locations on the first two detector layers, whereas the remaining vector is the sequel of the hit positions on the last two detector layers as illustrated in Fig.~\ref{Deflection_angle}.
\vspace*{-\baselineskip}
Upon these two vectors that suffice the computation of the angular deviation, the deflection angle of a muon crossing the detector layers denoted by $\theta$ is obtained by using the following expression~\cite{carlisle2012multiple, nugent2017multiple, poulson2019application}:
\begin{equation}
\theta=\arccos\left (\frac{\vec{v}_{1} \cdot \vec{v}_{2}}{\left|v_{1}\right|\left|v_{2}\right|}\right)
\end{equation}
Since the detector layers accept a significant number of muons, the average deflection angle at a particular energy is determined by averaging the previously calculated deflection angles over $N$ number of the non-absorbed/non-decayed muons as indicated in
\begin{equation}
\bar{\theta}=\frac{1}{N}\sum_{i=1}^{N}\theta_{i}
\end{equation}
which refers to a standard deviation as written in
\begin{equation}
\delta\theta=\sqrt{\frac{1}{N}\sum_{i=1}^{N}(\theta_{i}-\bar{\theta})^{2}}
\end{equation}
In light of the fact that the deflection angle is a consequence of a stochastic process, the standard deviation of the deflection angle is anticipated to be lessened with the intention of a better energy estimation. By harking back to the entire tomographic setup consisting of two separate sections (i.e. the top section and the bottom section), each of which includes three detector layers, the deflection angle determined for each non-absorbed/non-decayed muon in each section is first averaged over the number of sections, then the mean deflection angle of both sections is re-averaged over the number of the non-absorbed/non-decayed muons. Essentially, the following term outlines the average deflection angle of two separate hodoscopes specified by $x$ and $y$ at a certain energy value over $N$ number of the non-absorbed/non-decayed muons:
\begin{equation}
\bar{\theta}_{\frac{x+y}{2}}=\frac{1}{N}\sum_{i=1}^{N}\frac{\theta_{x, i}+\theta_{y, i}}{2}
\label{average}
\end{equation}
Accordingly, the resulting standard deviation dependent on the contributions from both the top hodoscope as well as the bottom hodoscope is formulated as noted in 
\begin{equation}
\delta\theta=\sqrt{\frac{1}{N}\sum_{i=1}^{N}\biggl(\frac{\theta_{x, i}+\theta_{y, i}}{2}-\bar{\theta}_{\frac{x+y}{2}}\biggr)^{2}}
\label{std}
\end{equation}
\section{Misclassification probabilities}
\label{sec:Mis}
By highlighting the probability that even fairly different muon energies routinely generate a substantial set of similar deflection angles, it is possible to foresee that the standard deviations at discrete energies are supposed to coincide in a considerably common sub-interval. In the first method, by assuming that two adjacent energy groups, i.e. labeled by $A$ and $B$ in a descending order, intersect in terms of the deflection angle by virtue of one standard deviation in one dimension, the misclassification probability is the ratio between the overlapping length $\bar{\theta}_{B}+\delta\theta_{B}-\bar{\theta}_{A}+\delta\theta_{A}$ and the entire length $\bar{\theta}_{A}+\delta\theta_{A}-\bar{\theta}_{B}+\delta\theta_{B}$ as follows
\begin{equation}
P_{\rm Linear}= \begin{cases} 
 \displaystyle\frac{\mbox{Overlapping length}}{\mbox{Entire length}}=\displaystyle\frac{\bar{\theta}_{B}+\delta\theta_{B}-\bar{\theta}_{A}+\delta\theta_{A}}{\bar{\theta}_{A}+\delta\theta_{A}-\bar{\theta}_{B}+\delta\theta_{B}} & \bar{\theta}_{B}+\delta\theta_{B} > \bar{\theta}_{A}-\delta\theta_{A} \\
 0 & \bar{\theta}_{B}+\delta\theta_{B} \leq \bar{\theta}_{A}-\delta\theta_{A}\\
 \end{cases}
\label{linear}
\end{equation}
In other words, two overlapping error bars defined by one standard deviation at two neighboring energy values are explicitly referred to the finite linear approximation in Eq. (\ref{linear}). 

In the second procedure, we consider that two positively defined modified Gaussian probability density functions (PDFs) represent the mean deflection angles and its standard deviations at kinetic energies $A$ and $B$, respectively, as follows
\begin{equation}
G'(\bar{\theta}_{A}, \delta\theta_{A}, \theta) = \frac{G(\bar{\theta}_{A}, \delta\theta_{A}, \theta)}{\displaystyle \int_{0}^{\infty}G(\bar{\theta}_{A}, \delta\theta_{A}, \theta)d\theta }
\label{GA}
\end{equation}
and
\begin{equation}
G'(\bar{\theta}_{B}, \delta\theta_{B}, \theta) = \frac{G(\bar{\theta}_{B}, \delta\theta_{B}, \theta)}{\displaystyle \int_{0}^{\infty}G(\bar{\theta}_{B}, \delta\theta_{B}, \theta)d\theta }
\label{GB}
\end{equation}
Then, the overlapping coefficient (OVL) of these two Gaussian PDFs might be obtained by calculating the intersection area as shown in
\begin{equation}
\mbox{OVL}=\mbox{Intersection area}=\displaystyle \int_{0}^{\infty} \min[G'(\bar{\theta}_{A}, \delta\theta_{A}, \theta) , G'(\bar{\theta}_{B}, \delta\theta_{B}, \theta)]d\theta 
\end{equation}
Thus, in the case where the intervals constitute a Gaussian distribution, the misclassification probability is determined by using the following expression:
 \begin{equation}
P_{\rm Gaussian}=\frac{\rm OVL}{\displaystyle \int_{0}^{\infty} G'(\bar{\theta}_{A}, \delta\theta_{A}, \theta)d\theta +\displaystyle \int_{0}^{\infty} G'(\bar{\theta}_{B}, \delta\theta_{B}, \theta)d\theta -{\rm OVL}}=\frac{\rm OVL}{\rm 2-OVL}
\end{equation}
\section{Simulation scheme}
\label{sec:Sim}
\begin{figure}[H]
\begin{center}
\includegraphics[width=6.7cm]{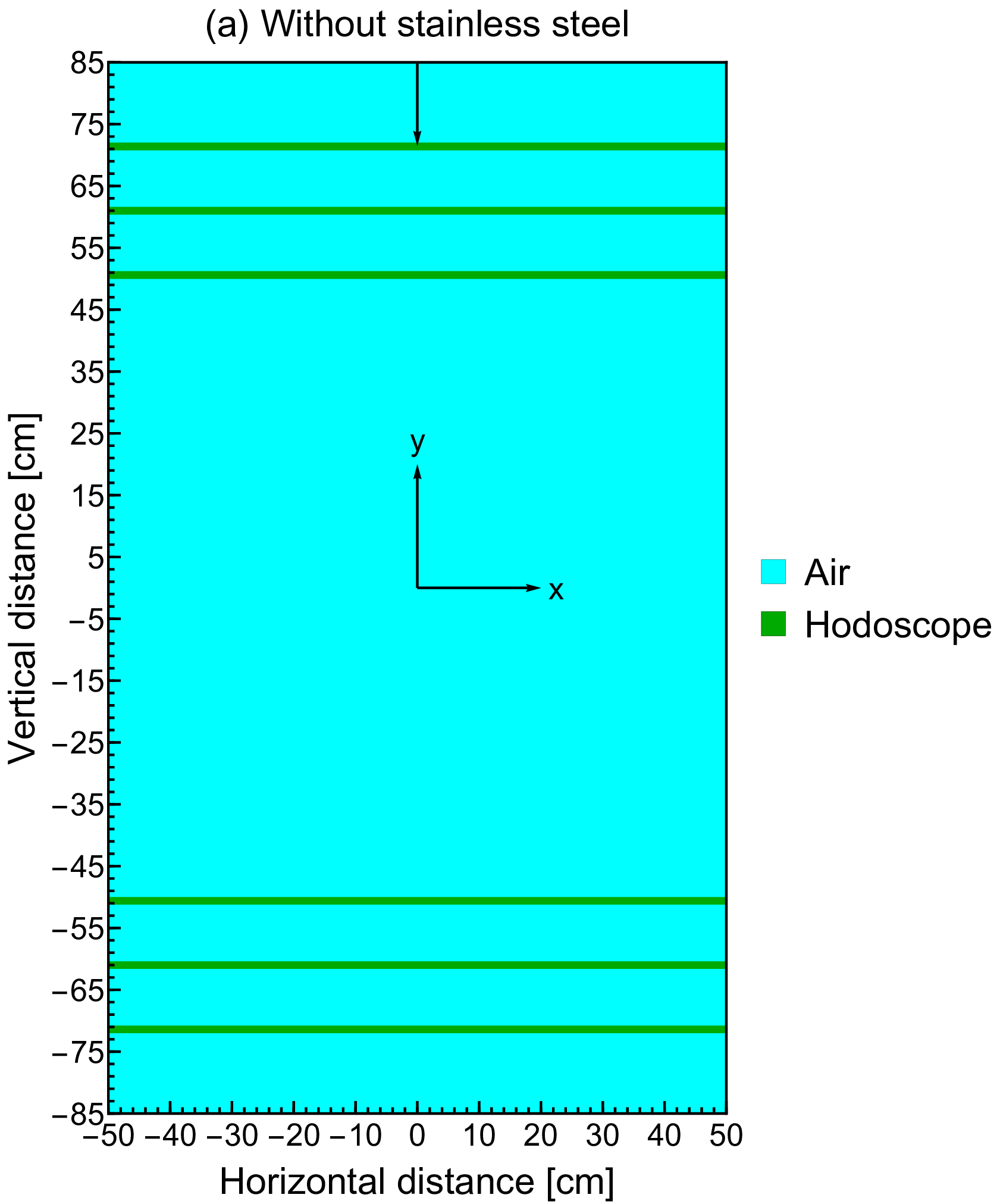}
\hskip 0.75 cm
\includegraphics[width=7.1cm]{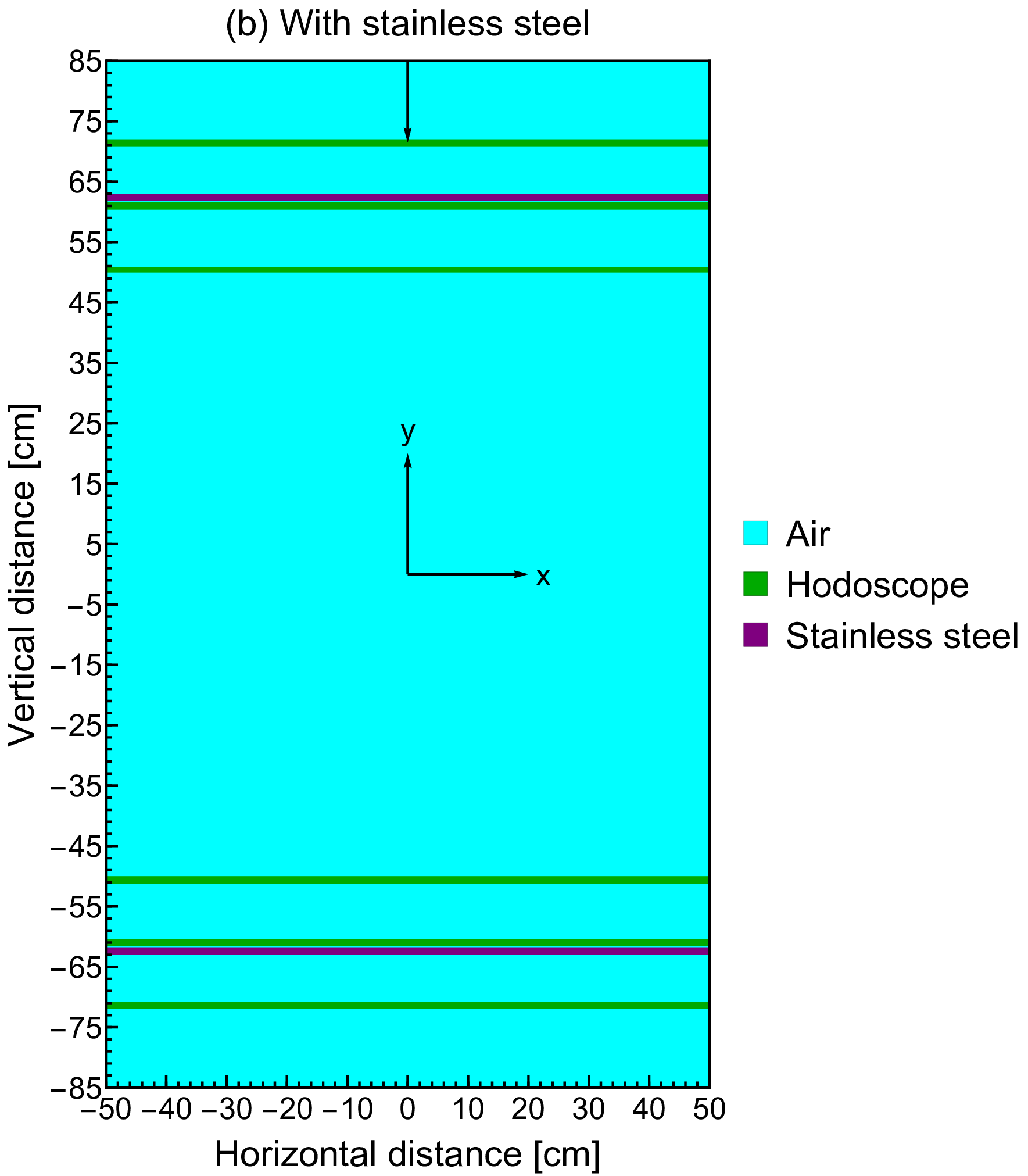}
\caption{Hodoscope schemes in the GEANT4 simulations (a) without stainless steel layers and (b) with stainless steel layers.}
\label{SimSch}
\end{center}
\end{figure}

The present study is conducted by means of the GEANT4 simulations, and we initially track the muon hit locations on the detector layers made out of polyvinyl toluene in order to calculate the average deflection angles as well as the corresponding standard deviations for the associated groups of kinetic energies. The simulation geometry for our present prototype without passive metallic layers in addition to the alternative hodoscope scheme containing stainless steel layers is depicted in Fig.~\ref{SimSch}. 
As described in Fig.~\ref{SimSch}(a), the dimensions of each the detector layer fabricated from polyvinyl toluene are $100\times0.4\times100$ $\rm cm^{3}$, and the inter-layer spacing at each section is 10 cm. In the proposed hodoscope setup, where we introduced stainless steel layers as illustrated in Fig.~\ref{SimSch}(b), the dimensions of each metallic layer are $100\times0.4\times100$ $\rm cm^{3}$, i.e. identical to those of the detector layers. In both the tomographic schemes, the gap between the top hodoscope and the bottom hodoscope is 100 cm into the bargain.

A central mono-directional beam with the uniform energy distribution that is generated at y=85 cm via G4ParticleGun is employed, and the generated muons are crossing in the vertically downward direction, i.e. from the top edge of the simulation box through the bottom edge as indicated with the black arrows in Fig.~\ref{SimSch}. Since the current aperture of the entire detection geometry commonly only accepts the narrow angles apart from the very rare entries around the corners, this beam setup is considered significantly reliable by reminding that the distribution of the incident angle ($\alpha$) approximately corresponds to $\cos^2\alpha$ for an interval between $-\pi/2$ and $\pi/2$~\cite{yanez2021method}. We favor a uniform energy distribution lying on an energy interval between 0.25 and 8 GeV so as to achieve the absorption minimization especially in the case of stainless steel layers besides the numerical accuracy optimization~\cite{anghel2015plastic}. We partition the entire energy interval bounded by 0.25 and 8 GeV into three energy groups such that the first energy group consists of the energy values between 0.25 and 0.75 GeV, whereas the second energy group is composed of the muon energies between 0.75 and 3.75 GeV, and the muon energies between 3.75 and 8 GeV finally constitute the third energy group. 
\vspace*{-1.4\baselineskip} 
For each of the three energy groups, the number of the simulated muons is 10$^{4}$. All the materials in the current study are defined in accordance with the GEANT4/NIST material database. The reference physics list used in these simulations is FTFP$\_$BERT. 
The muon tracking is sustained by G4Step, and the registered hit positions are post-processed with the assistance of a Python script where the deflection angle is first determined for every single non-absorbed/non-decayed muon of a given energy group at each hodoscope. Finally, we conclude the post-processing stage by applying Eqs.~(\ref{average}) and (\ref{std}) on the datasets of deflection angle attained through the top hodoscope and the bottom hodoscope, thereby obtaining the average deflection angle and the corresponding standard deviation.
\section{Simulation outcomes}
\label{sec:Out}
\begin{figure}[H]
\begin{center}
\includegraphics[width=7.3cm]{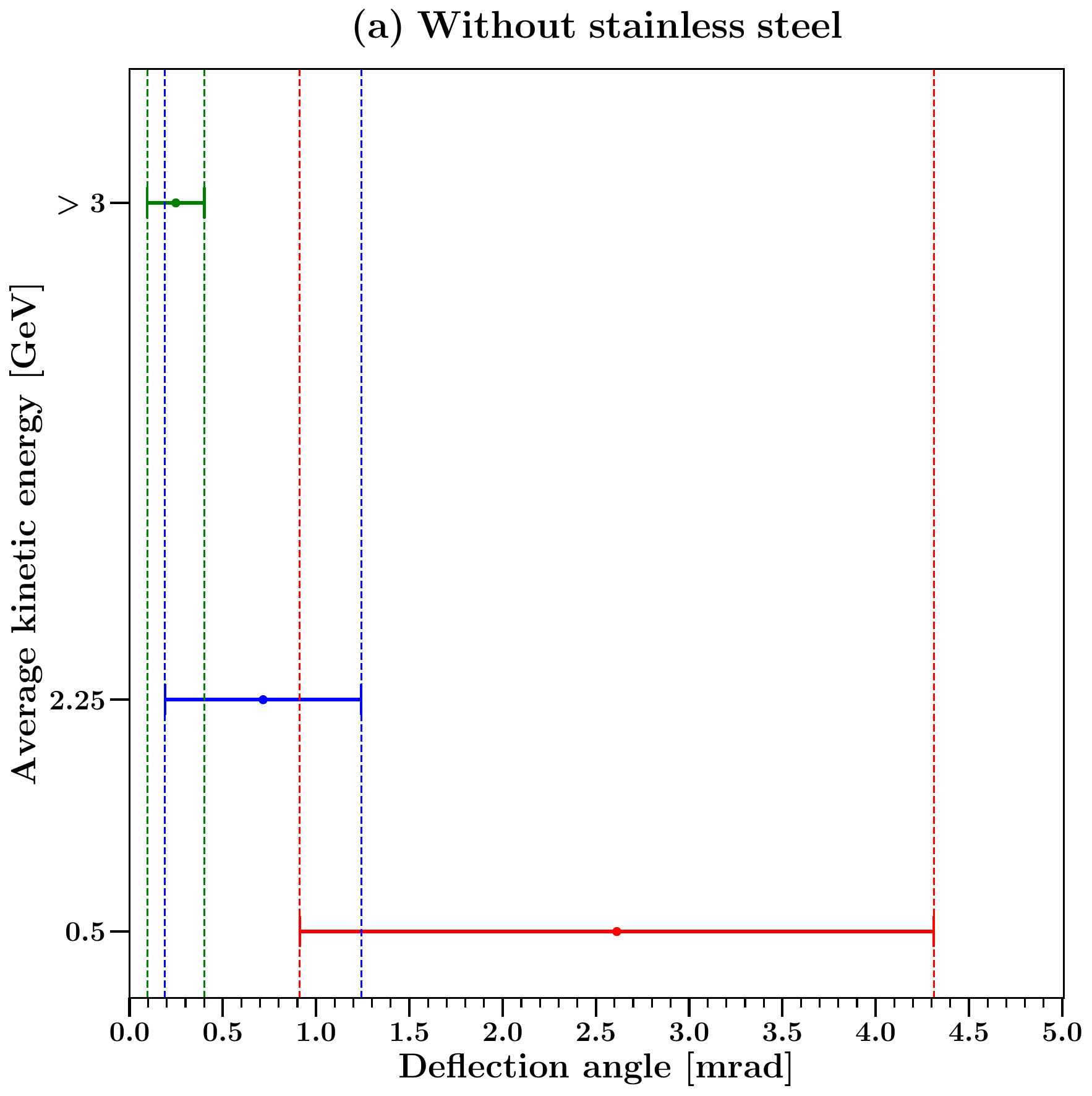}
\includegraphics[width=7.3cm]{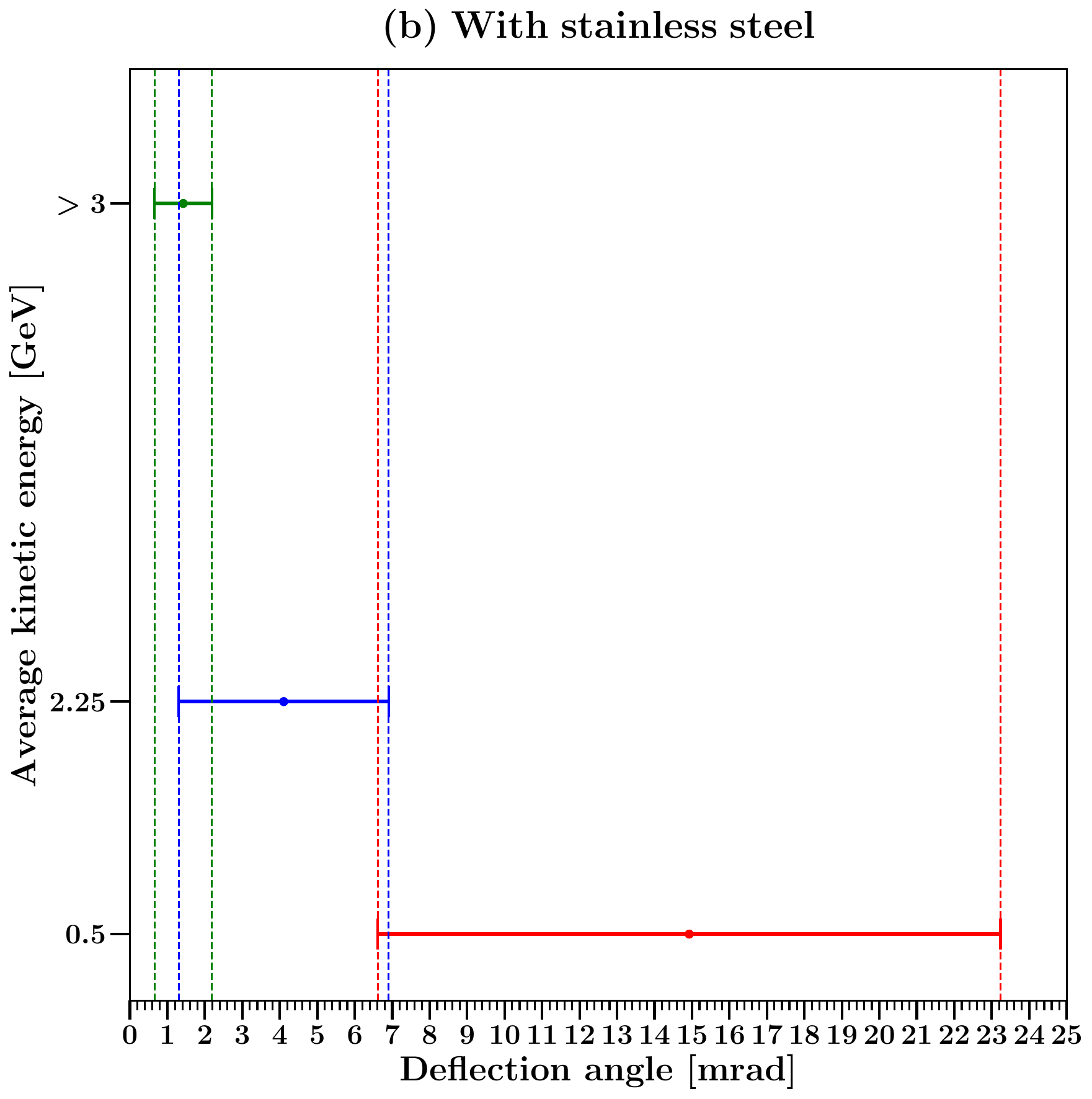}
\caption{Comparison of linear overlaps over a three-group energy categorization by using a finite linear approximation for the simulation cases (a) without stainless steel layers and (b) with stainless steel layers.}
\label{linearoverlap}
\end{center}
\end{figure}

\begin{figure}[H]
\begin{center}
\textbf{Without stainless steel}\\
\includegraphics[width=6cm]{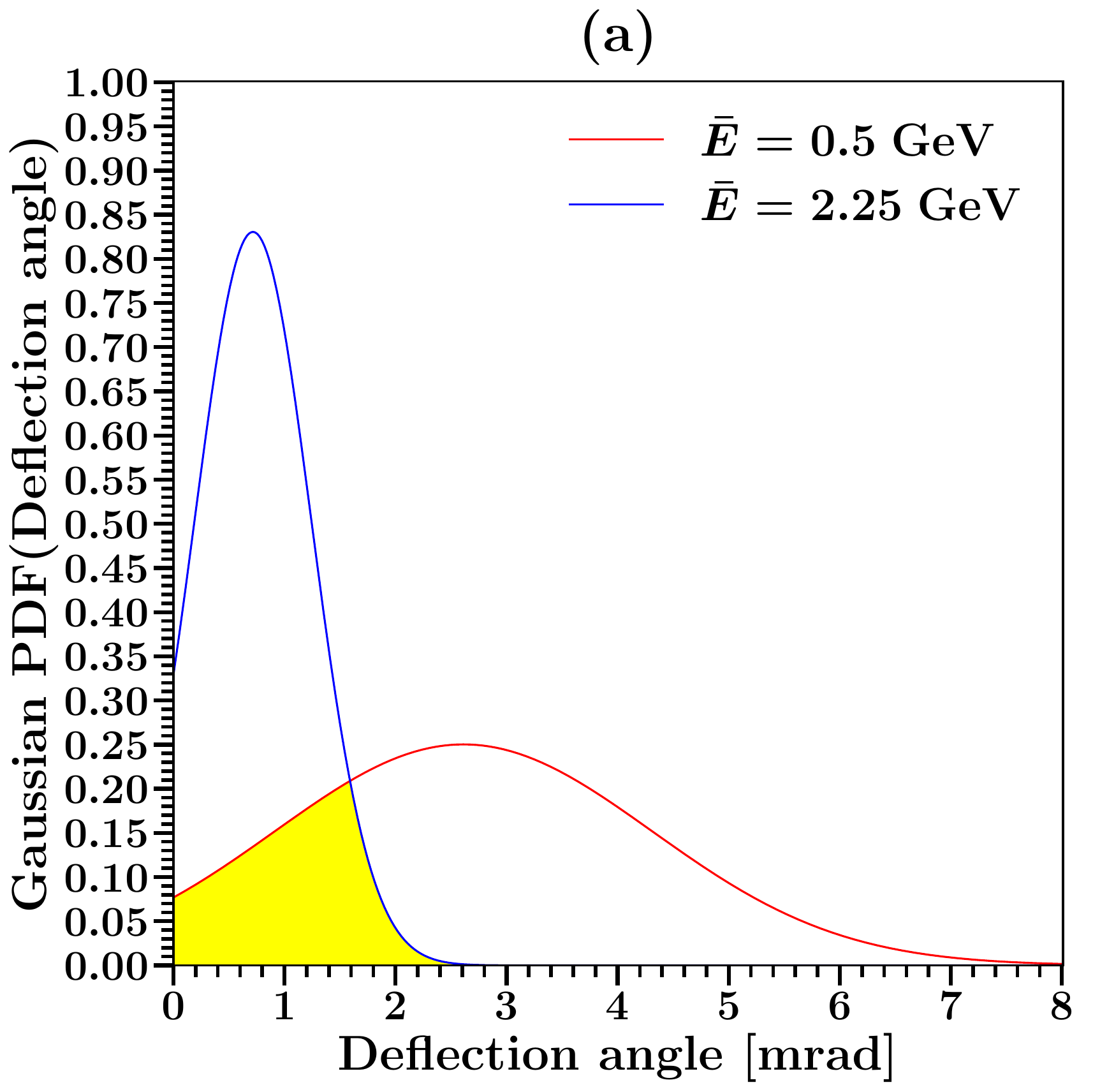}
\includegraphics[width=6cm]{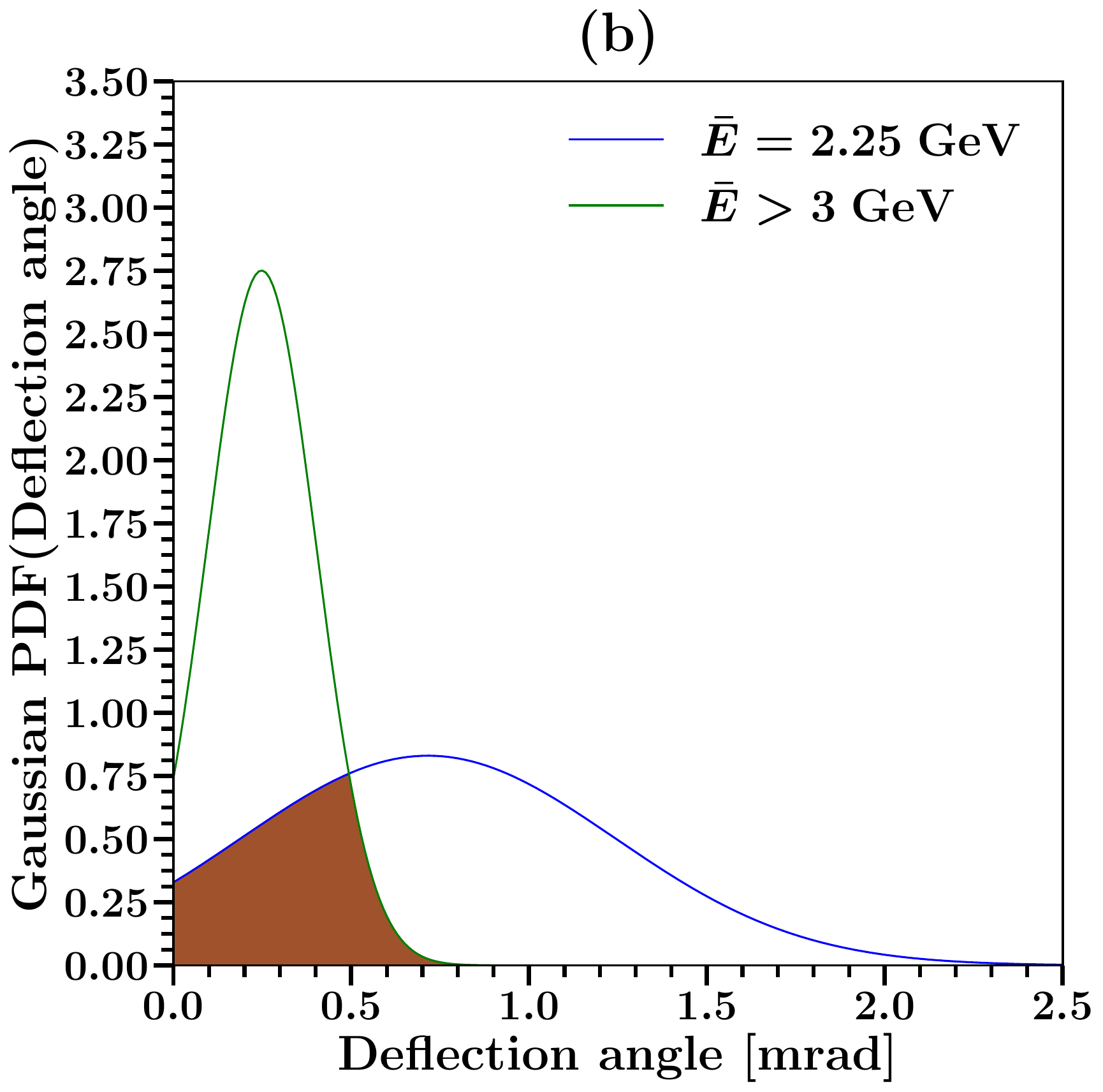}\\
\textbf{With stainless steel}\\
\includegraphics[width=6cm]{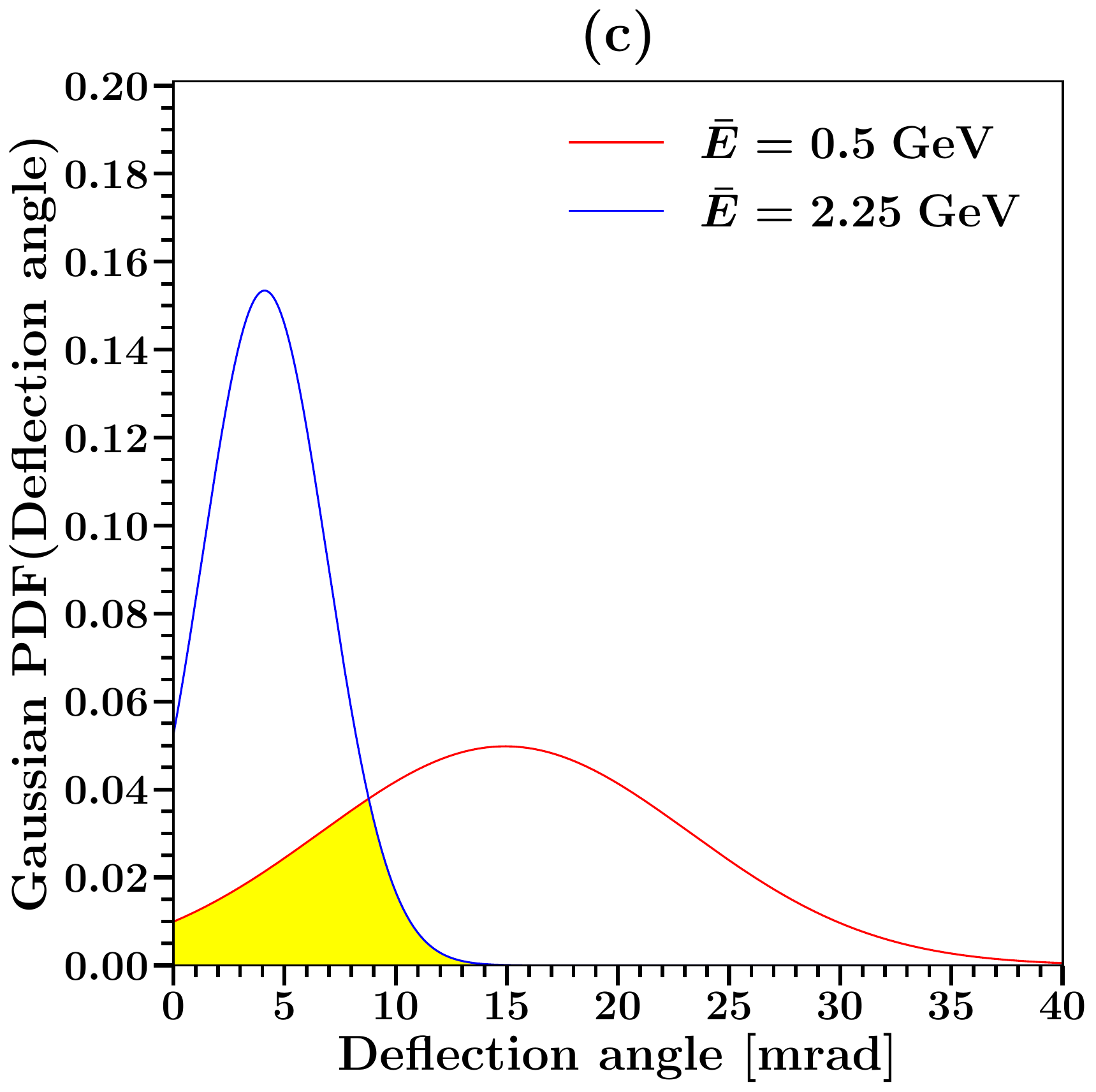}
\includegraphics[width=6cm]{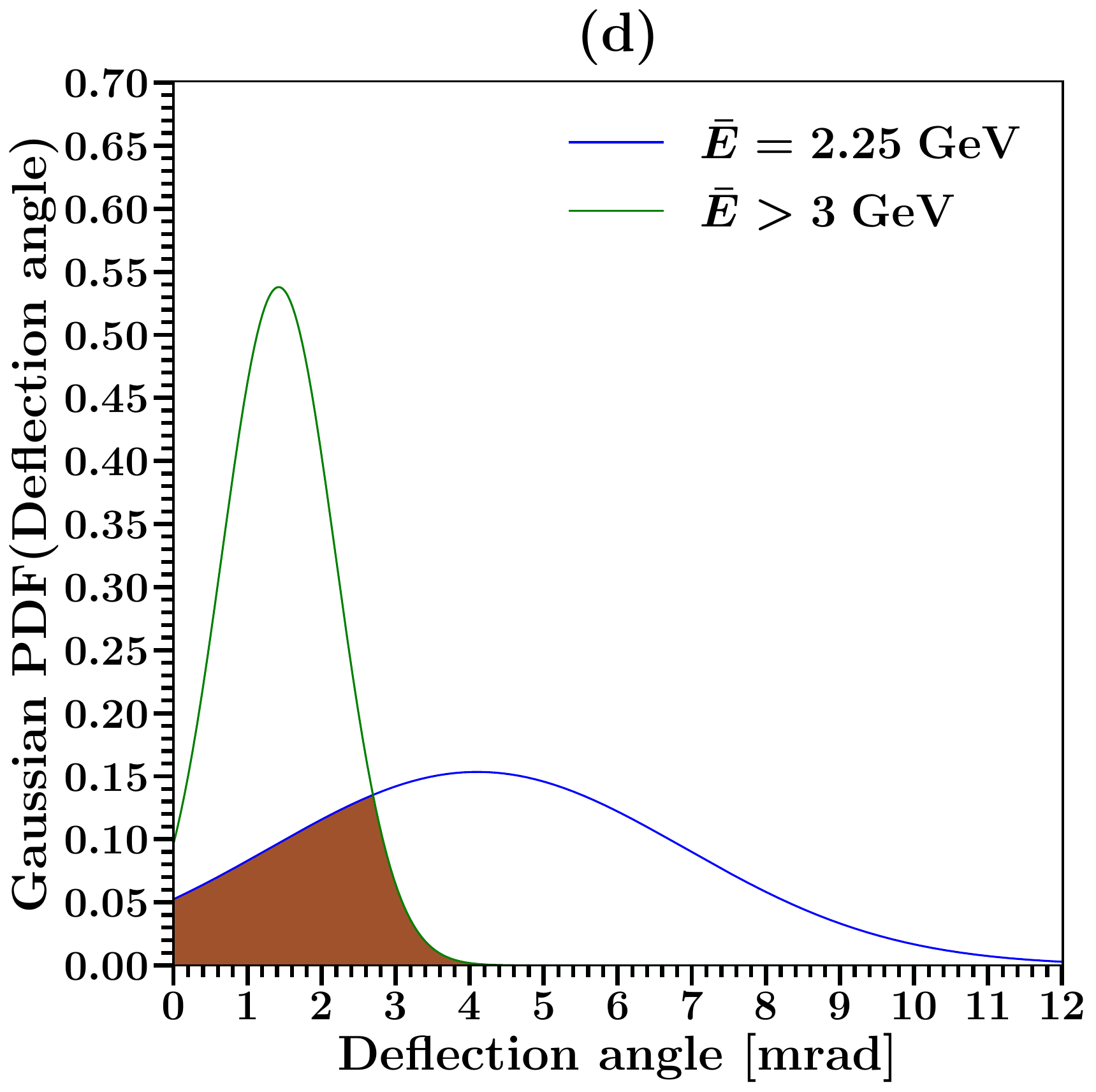}
\caption{Contrast between the hodoscope schemes (a)-(b) without stainless steel layers and (c)-(d) with stainless steel layers over a three-group energy classification by means of positively defined modified Gaussian distributions.}
\label{Gaussianoverlap}
\end{center}
\end{figure}

Our GEANT4 simulations initially yield the average deflection angles as well as the associated standard deviations for both our current prototype and our proposed new scheme as tabulated in Table~\ref{avgangle}. By contrasting the present prototype with the alternative hodoscope in agreement with the average deflection angles in Table~\ref{avgangle}, the first positive influence of stainless steel layers in terms of the detection performance is already revealed. To better elucidate, by reminding the existing detector accuracy of 1 mrad, all the average deflection angles generated by the proposed hodoscope containing stainless steel layers exceed the angular value of 1 mrad for the three-group energy structure; on the other hand, the average deflection angles obtained through the present configuration remains below this detector accuracy in two among three energy groups, i.e. $\bar{E}=2.25$ GeV and $\bar{E}>3$ GeV.
\vskip -0.35cm 
\begin{table}[H]
\begin{footnotesize}
\begin{center}
\caption{Average deflection angles and standard deviations obtained through a three-group energy structure for the cases without stainless steel layers and with stainless steel layers, respectively.}
\vspace*{-0.75\baselineskip} 
\resizebox{\textwidth}{!}{\begin{tabular}{*5c}
\toprule
\toprule
& & \textbf{Without stainless steel} &\\
\midrule
Energy interval [GeV] & $\bar{E}$ [GeV] & $\bar{\theta}_{\rm Top} \pm\delta\theta$ [mrad] & $\bar{\theta}_{\rm Bottom} \pm\delta\theta$ [mrad] & $\bar{\theta}_{\rm \frac{Top+Bottom}{2}} \pm\delta\theta$ [mrad] \\
\midrule
0.25 - 0.75 & 0.5 & 2.590$\pm$2.107 & 2.632$\pm$2.463 & 2.612$\pm$1.700\\
0.75 - 3.75 & 2.25 & 0.712$\pm$0.657 & 0.719$\pm$0.686 &0.716$\pm$0.526\\
3.75 - 8 & $>3$ & 0.248$\pm$0.217 & 0.249$\pm$0.204 & 0.248$\pm$0.153\\
\midrule
& & \textbf{With stainless steel} &\\
\midrule
Energy interval [GeV] & $\bar{E}$ [GeV] & $\bar{\theta}_{\rm Top} \pm\delta\theta$ [mrad] & $\bar{\theta}_{\rm Bottom} \pm\delta\theta$ [mrad] & $\bar{\theta}_{\rm \frac{Top+Bottom}{2}} \pm\delta\theta$ [mrad] \\
\midrule
0.25 - 0.75 & 0.5 & 14.797$\pm$10.915 & 15.055$\pm$10.871&14.925$\pm$8.310\\
0.75 - 3.75 & 2.25 & 4.080$\pm$3.459 & 4.128$\pm$3.559 & 4.104$\pm$2.800\\
3.75 - 8 & $>3$ & 1.418$\pm$1.051 & 1.427$\pm$1.034 & 1.422$\pm$0.766\\
\bottomrule
\bottomrule
\label{avgangle}
\end{tabular}}
\end{center}
\end{footnotesize}
\end{table}
\vspace*{-1.8\baselineskip} 
On top of the augmented average deflection angles that lead to a better detection efficiency, the presence of stainless steel layers is partially expected to result in the diminished uncertainty. To verify this hypothesis over our GEANT4 simulations, the linear overlap based on the standard deviations is shown in Fig.~\ref{linearoverlap}, and we observe that the uncertainty in the deflection angle is apparently decreased in the case of stainless layers compared to the present hodoscope scheme without any passive metallic layers.
\vspace*{-0.75\baselineskip} 
In order to further evaluate the effect of stainless steel layers, the areal overlap of the positively defined modified Gaussian distributions as defined in Eqs.~(\ref{GA}) and (\ref{GB}) is displayed in Fig.~\ref{Gaussianoverlap}, and we face a similar reduction in the angular uncertainty for the reason of the areal contraction. 
\vspace*{-0.75\baselineskip} 
In the long run, the misclassification probabilities calculated by means of these methodologies are listed in Table~\ref{Mis}, and it is numerically exhibited that the proposed hodocope scheme with stainless steel layers leads to the remarkably lower misclassification probabilities in comparison with the present tomographic setup without stainless steel layers.
\vskip -0.675cm 
\begin{table}[H]
\begin{footnotesize}
\begin{center}
\parbox{0.75\textwidth}{\caption{Misclassification probabilities for the cases without stainless steel layers and with stainless steel layers, respectively.}}
\label{Mis}
\resizebox{0.75\textwidth}{!}{\begin{tabular}{*5c}
\toprule
\toprule
&~~~~~~~~~~~~~~~~\textbf{Without stainless steel} &\\
\midrule
$\bar{E}$ pairs [GeV] & $\bar{\theta}_{\rm \frac{Top+Bottom}{2}} \pm\delta\theta$ pairs [mrad] & OVL & $P_{\rm Gaussian}$ & $P_{\rm Linear}$ \\
\midrule
0.5 - 2.25 & 2.612$\pm$1.700 - 0.716$\pm$0.526 & 0.278 & 0.161 & 0.080 \\
2.25 - $>3$ & 0.716$\pm$0.526 - 0.248$\pm$0.153 & 0.330 & 0.197 & 0.180\\
\midrule
&~~~~~~~~~~~~~~~~\textbf{With stainless steel} &\\
\midrule
$\bar{E}$ pairs [GeV] & $\bar{\theta}_{\rm \frac{Top+Bottom}{2}} \pm\delta\theta$ pairs [mrad] & OVL & $P_{\rm Gaussian}$ & $P_{\rm Linear}$ \\
\midrule
0.5 - 2.25 & 14.925$\pm$8.310 - 4.104$\pm$2.800 & 0.251 & 0.144 & 0.013\\
2.25 - $>3$ & 4.104$\pm$2.800 - 1.422$\pm$0.766 & 0.304 & 0.179 & 0.141\\
\bottomrule
\bottomrule
\end{tabular}}
\end{center}
\end{footnotesize}
\end{table}
\vspace*{-1\baselineskip} 
\section{Conclusion}
In this study, we investigate an alternative tomographic setup by introducing stainless steel layers into the bottom section and the top section with the aim of the uncertainty reduction in the deflection angle. Upon our GEANT4 simulations combined with two distinct forms of the misclassification probabilities, we experience that the stainless steel layers slightly yield more favorable angular values with the diminished uncertainty as opposed to our current prototype apart from the increased average deflection angles that fulfill the present detector accuracy. Within the perspective of an indirect energy classification dependent on the deflection angle, our simulation outcomes qualitatively and quantitatively indicate that the presence of stainless steel layers might be an inexpensive upgrade in the potential applications of the muon scattering tomography. 
\label{sec:Con}
\bibliographystyle{ieeetr}
\bibliography{effect_of_passive_layers.bib}

\begin{thebibliography}{10}

\bibitem{Checchia_2016}
P.~Checchia, ``Review of possible applications of cosmic muon tomography,''
  {\em J. Instrum.}, vol.~11, no.~12, p.~C12072, 2016.

\bibitem{procureur2018muon}
S.~Procureur, ``{Muon imaging: Principles, technologies and applications},''
  {\em Nucl. Instr. Meth. A}, vol.~878, p.~169, 2018.

\bibitem{bonechi2020atmospheric}
L.~Bonechi {\em et~al.}, ``Atmospheric muons as an imaging tool,'' {\em Rev.
  Phys.}, vol.~5, p.~100038, 2020.

\bibitem{lee2017characteristics}
C.~H. Lee {\em et~al.}, ``Characteristics of plastic scintillators fabricated
  by a polymerization reaction,'' {\em Nucl. Eng. Technol.}, vol.~49, no.~3,
  p.~592, 2017.

\bibitem{borozdin2003radiographic}
K.~Borozdin {\em et~al.}, ``Radiographic imaging with cosmic-ray muons,'' {\em
  Nature}, vol.~422, no.~6929, p.~277, 2003.

\bibitem{thomay2016passive}
C.~Thomay {\em et~al.}, ``{Passive 3D imaging of nuclear waste containers with
  muon scattering tomography},'' {\em J. Instrum.}, vol.~11, no.~03, p.~P03008,
  2016.

\bibitem{frazao2019high}
L.~Fraz{\~a}o {\em et~al.}, ``High-resolution imaging of nuclear waste
  containers with muon scattering tomography,'' {\em J. Instrum.}, vol.~14,
  no.~08, p.~P08005, 2019.

\bibitem{anghel2015plastic}
V.~Anghel {\em et~al.}, ``A plastic scintillator-based muon tomography system
  with an integrated muon spectrometer,'' {\em Nucl. Instr. Meth. A}, vol.~798,
  p.~12, 2015.

\bibitem{georgadze2021method}
A.~Georgadze {\em et~al.}, ``Method and apparatus for detection and/or
  identification of materials and of articles using charged particles,'' Jan.~7
  2021.
\newblock US Patent App. 16/977,293.

\bibitem{agostinelli2003geant4}
S.~Agostinelli {\em et~al.}, ``{GEANT4 - a simulation toolkit},'' {\em Nucl.
  Instr. Meth. A}, vol.~506, no.~3, p.~250, 2003.

\bibitem{carlisle2012multiple}
T.~Carlisle {\em et~al.}, ``{Multiple Scattering Measurements in the MICE
  Experiment},'' tech. rep., Fermi National Accelerator Lab.(FNAL), Batavia, IL
  (United States), 2012.

\bibitem{nugent2017multiple}
J.~C. Nugent, {\em {Multiple Coulomb scattering in the MICE experiment}}.
\newblock PhD thesis, University of Glasgow, 2017.

\bibitem{poulson2019application}
D.~Poulson {\em et~al.}, ``Application of muon tomography to fuel cask
  monitoring,'' {\em Phil. Trans. R. Soc. A}, vol.~377, no.~2137, p.~20180052,
  2019.

\bibitem{yanez2021method}
B.~O. Y{\'a}{\~n}ez and A.~A. Aguilar-Arevalo, ``A method to measure the
  integral vertical intensity and angular distribution of atmospheric muons
  with a stationary plastic scintillator bar detector,'' {\em Nucl. Instr.
  Meth. A}, vol.~987, p.~164870, 2021.

\end{thebibliography}
\end{document}